\UseRawInputEncoding
\documentclass[twocolumn]
{aastex631}

\usepackage{bm} 
\usepackage{enumitem}
\usepackage{amsmath}
\usepackage{rotating}

\newcommand\edits[1]{#1}
\newcommand\nicole[1]{}
\newcommand\newedits[1]{#1}

\begin{document}

\title{ODIN: Star Formation Histories Reveal Formative Starbursts Experienced 
\\ by Lyman Alpha Emitting Galaxies at Cosmic Noon}


\author[0000-0002-9811-2443]{Nicole M. Firestone}
\affiliation{Department of Physics and Astronomy, Rutgers, the State University of New Jersey, Piscataway, NJ 08854, USA}

\author[0000-0003-1530-8713]{Eric Gawiser}
\affiliation{Department of Physics and Astronomy, Rutgers, the State University of New Jersey, Piscataway, NJ 08854, USA}
\affiliation{School of Natural Sciences, Institute for Advanced Study, Princeton, NJ 08540, USA}

\author[0000-0001-9298-3523]{Kartheik G. Iyer}
\affiliation{Columbia Astrophysics Laboratory, Columbia University, 550 West 120th Street, New York, NY 10027, USA}

\author[0000-0003-3004-9596]{Kyoung-Soo Lee}
\affiliation{Department of Physics and Astronomy, Purdue University, 525 Northwestern Ave., West Lafayette, IN 47906, USA}

\author[0000-0002-9176-7252]{Vandana Ramakrishnan}
\affiliation{Department of Physics and Astronomy, Purdue University, 525 Northwestern Ave., West Lafayette, IN 47906, USA}

\author[0000-0001-5567-1301]{Francisco Valdes}
\affiliation{NSF’s National Optical-Infrared Astronomy Research Laboratory, 950 N. Cherry Ave., Tucson, AZ 85719, USA}

\author[0000-0001-9521-6397]{Changbom Park}
\affiliation{Korea Institute for Advanced Study, 85 Hoegi-ro, Dongdaemun-gu, Seoul 02455, Republic of Korea}

\author[0000-0003-3078-2763]{Yujin Yang}
\affiliation{Korea Astronomy and Space Science Institute, 776 Daedeokdae-ro, Yuseong-gu, Daejeon 34055, Republic of Korea}


\author[0000-0002-8630-6435]{Anahita Alavi}
\affiliation{IPAC, Mail Code 314-6, California Institute of Technology, 1200 E. California Blvd., Pasadena, CA 91125, USA}

\author[0000-0002-1328-0211]{Robin Ciardullo}
\affiliation{Department of Astronomy \& Astrophysics, The Pennsylvania
State University, University Park, PA 16802, USA}
\affiliation{Institute for Gravitation and the Cosmos, The Pennsylvania
State University, University Park, PA 16802, USA}






\author[0000-0001-9440-8872]{Norman Grogin}
\affiliation{Space Telescope Science Institute, 3700 San Martin Drive, Baltimore, MD 21218, USA}

\author[0000-0001-6842-2371]{Caryl Gronwall}
\affiliation{Department of Astronomy \& Astrophysics, The Pennsylvania
State University, University Park, PA 16802, USA}
\affiliation{Institute for Gravitation and the Cosmos, The Pennsylvania
State University, University Park, PA 16802, USA}

\author[0000-0002-4902-0075]{Lucia Guaita}
\affiliation{Universidad Andres Bello, Facultad de Ciencias Exactas, Departamento de Fisica y Astronomia, Instituto de Astrofisica, Fernandez Concha 700, Las Condes, Santiago RM, Chile}
\affiliation{Millennium Nucleus for Galaxies}


\author[0000-0001-9991-8222]{Sungryong Hong}
\affiliation{Korea Astronomy and Space Science Institute, 776 Daedeokdae-ro, Yuseong-gu, Daejeon 34055, Republic of Korea}

\author[0000-0003-3428-7612]{Ho Seong Hwang}
\affiliation{Department of Physics and Astronomy, Seoul National University, 1 Gwanak-ro, Gwanak-gu, Seoul 08826, Republic of Korea}
\affiliation{SNU Astronomy Research Center, Seoul National University, 1 Gwanak-ro, Gwanak-gu, Seoul 08826, Republic of Korea}
\affiliation{Australian Astronomical Optics - Macquarie University, 105 Delhi Road, North Ryde, NSW 2113, Australia}

\author[0009-0003-9748-4194]{Sang Hyeok Im}
\affiliation{Department of Physics and Astronomy, Seoul National University, 1 Gwanak-ro, Gwanak-gu, Seoul 08826, Republic of Korea}

\author[0000-0002-2770-808X]{Woong-Seob Jeong}
\affiliation{Korea Astronomy and Space Science Institute, 776 Daedeokdae-ro, Yuseong-gu, Daejeon 34055, Republic of Korea}


\author{Seongjae Kim}
\affiliation{Korea Astronomy and Space Science Institute, 776 Daedeokdae-ro, Yuseong-gu, Daejeon 34055, Republic of Korea}
\affiliation{University of Science and Technology, 217 Gajeong-ro, Yuseong District, Daejeon, Republic of Korea}

\author[0000-0002-6610-2048]{Anton M. Koekemoer}
\affiliation{Space Telescope Science Institute, 3700 San Martin Drive,
Baltimore, MD 21218, USA}

\author{Ankit Kumar}
\affiliation{Universidad Andres Bello, Facultad de Ciencias Exactas, Departamento de Fisica y Astronomia, Instituto de Astrofisica, Fernandez Concha 700, Las Condes, Santiago RM, Chile}

\author[0000-0002-6810-1778]{Jaehyun Lee}
\affiliation{Korea Astronomy and Space Science Institute, 776 Daedeokdae-ro, Yuseong-gu, Daejeon 34055, Republic of Korea}

\author[0000-0001-7166-6035]{Vihang Mehta}
\affiliation{IPAC, Mail Code 314-6, California Institute of Technology, 1200 E. California Blvd., Pasadena, CA 91125, USA}

\author[0000-0002-0905-342X]{Gautam Nagaraj}
\affiliation{Laboratoire d'Astrophysique, EPFL, 1015 Lausanne, Switzerland}

\author{Julie Nantais}
\affiliation{Universidad Andres Bello, Facultad de Ciencias Exactas, Departamento de Fisica y Astronomia, Instituto de Astrofisica, Fernandez Concha 700, Las Condes, Santiago RM, Chile}




\author[0000-0002-0604-654X]{Laura Prichard}
\affiliation{Space Telescope Science Institute, 3700 San Martin Drive, Baltimore, MD 21218, USA}




\author[0000-0002-9946-4731]{Marc Rafelski}
\affiliation{Space Telescope Science Institute, 3700 San Martin Drive, Baltimore, MD 21218, USA}
\affiliation{Department of Physics and Astronomy, Johns Hopkins University, Baltimore, MD 21218, USA}

\author[0000-0002-4362-4070]{Hyunmi Song}
\affiliation{Department of Astronomy and Space Science, Chungnam National University, 99 Daehak-ro, Yuseong-gu, Daejeon, 34134, Republic of Korea}

\author[0000-0003-3759-8707]{Ben Sunnquist}
\affiliation{Space Telescope Science Institute, 3700 San Martin Drive, Baltimore, MD 21218, USA}

\author[0000-0002-7064-5424]{Harry I. Teplitz}
\affiliation{IPAC, Mail Code 314-6, California Institute of Technology, 1200 E. California Blvd., Pasadena, CA 91125, USA}

\author[0000-0002-9373-3865]{Xin Wang}
\affiliation{School of Astronomy and Space Science, University of Chinese Academy of Sciences (UCAS), Beijing 100049, China}
\affiliation{National Astronomical Observatories, Chinese Academy of Sciences, Beijing 100101, China}
\affiliation{Institute for Frontiers in Astronomy and Astrophysics, Beijing Normal University, Beijing 102206, China}




\begin{abstract}


In this work, we test the \newedits{frequent} assumption that Lyman Alpha Emitting galaxies (LAEs) are experiencing their first major burst of star formation at the time of observation. To this end, we identify 74 LAEs from the ODIN Survey with rest-UV-through-NIR photometry from UVCANDELS. For each LAE, we perform non-parametric star formation history (SFH) reconstruction using the Dense Basis Gaussian process-based method of spectral energy distribution fitting. We find that a strong majority (67\%) of our LAE SFHs align with the \newedits{frequently assumed} archetype of a first major star formation burst, with at most modest star formation rates (SFRs) in the past. However, the rest of our LAE SFHs have significant amounts of star formation in the past, with 28\% exhibiting earlier bursts of star formation with the ongoing burst having the highest SFR (\textit{dominant bursts}), and the final 5\% having experienced their highest SFR in the past (\textit{non-dominant bursts}). Combining the SFHs indicating \textit{first} and \textit{dominant bursts}, $\sim$95\% of LAEs are experiencing their largest burst yet-- a \textit{formative} burst. We also find that the fraction of total stellar mass created in the last 200\,Myr is $\sim$1.3 times higher in LAEs than in mass-matched Lyman Break Galaxy (LBG) samples, and that a majority of LBGs are experiencing \textit{dominant bursts}, reaffirming that LAEs differ from other star forming galaxies. Overall, our results suggest that multiple evolutionary paths can produce galaxies with strong observed Ly$\alpha$ emission. 

\end{abstract}



\section{Introduction}\label{sec:intro}


Lyman Alpha Emitting (LAE) galaxies are widely used to study both galaxy formation and cosmology because they provide the most readily accessible observational probe of low-mass, high-redshift galaxies. \edits{\newedits{Historically}, LAEs \newedits{have often been broadly characterized} as young, low-mass, low-dust galaxies experiencing their first major burst of star formation at the time \edits{of observation} (\citealp{Partridge1967}; \citealp[see also ][]{maier2003constraints, rhoads2003spectroscopic, venemans2005properties, thommes2005expected, arrabal2020differences}). The resulting strong Ly$\alpha$ emission at a rest-frame wavelength of 121.6\,nm allows LAEs to serve as highly identifiable probes of the high-redshift universe.} In both blind spectroscopic surveys and narrowband imaging surveys, the primary criterion for classifying an object as an LAE is that its observed rest-frame Ly$\alpha$ equivalent width \edits{(EW)} must \edits{reach a certain threshold (typically taken to be 20 {\AA} \citep{Ouchi_2020})}. This carries an implication of active star formation from hot young O-type and B-type stars \citep[e.g.,][]{kunth1998hst, hui1997equation} compared to the ultraviolet continuum emission from O-type, B-type, and A-type stars (in cases where Ly$\alpha$ emission is not driven by \edits{active galactic nuclei}). However, \edits{Ly$\alpha$ EW} does not reveal any information about \edits{the} level of star formation in the context of a galaxy's lifetime or the underlying radiative transfer scenarios that may generate such a high observed equivalent width. \edits{The question of} whether or not LAEs \newedits{generally} assemble stellar mass in a congruent and unique way \edits{remains open}. For example, \newedits{what fraction of} LAEs \newedits{are} young galaxies in their first burst of star formation at the time of observation, as the \citet{Partridge1967} \nicole{removed ``conventional'' here} model of LAEs implies? Can an LAE only be an LAE once or can a galaxy become an LAE several times throughout its lifetime? \edits{Answers to these questions will provide insight into the complex evolutionary paths and intergalactic conditions that make LAEs such profound beacons of the high-$z$ universe.} In order to \edits{begin to} answer these questions, we turn to spectral energy distributions (SEDs).


With the advent of wide-field multi-wavelength imaging surveys, we have been able to create SEDs \edits{of high redshift galaxies} that span rest-frame wavelengths ranging from \edits{the} ultraviolet (UV) to near-infrared (NIR) \citep[e.g.,][]{papovich2001stellar, shapley2011physical}. This has allowed us to fit \edits{these} SEDs to spectral templates with well-understood features to constrain the physical properties of galaxies. \citet{gawiser2007lyalpha} used this technique to conclude that \edits{a typical LAE} at $z$ = 3.1 \edits{has} an average stellar age of 20\,Myr, a low stellar mass of $\sim10^9\,M_\odot$, a moderate star formation rate (SFR) of 2$\,M_\odot$~yr$^{-1}$, and negligible dust extinction of $A_V < 0.2$. \citet{guaita2011lyalpha} conducted a similar analysis at $z$ = 2.4 and also found that LAEs are typically low-mass galaxies with moderate dust content of $E(B - V)$ = 0.22. Their results suggested that $z$ = 2.4 LAEs are dominated by young stars with a median stellar age of 10\,Myr, implying that LAEs are undergoing an intense burst of star formation at the time of observation. Further, \citet{vargas2014stack} used flux stacked SEDs from CANDELS to determine that LAEs at $z$ = 2.1 have a median stellar mass of $3 \times 10^8$\,$M_\odot$, \edits{a median} age of 100\,Myr, and \edits{a typical extinction of} $E(B - V)$ of 0.12. \edits{Comparably,} \citet{acquaviva2012curious} used stacked SEDs of $z$ = 2.1 and $z$ = 3.1 LAEs to investigate the typical physical properties of LAEs at each of these redshifts. They found that 
it is unlikely for $z$ = 3.1 LAEs to evolve into $z$ = 2.1 LAEs\edits{, suggesting that there is a discrepancy between the expected evolutionary path of LAEs from clustering analysis and SED results}. \citet{kusakabe2018stellar} measured stacked LAE properties over four different fields using SED fitting and found a mean stellar mass of $10.2 \times 10^8$\,$M_\odot$, a mean age of $380$\,Myr, and a mean SFR of 3.4\,$M_\odot$~yr$^{-1}$. While these studies affirm the characterization of LAEs as generally young, low-mass, low-dust galaxies, there are notable discrepancies in their stacked measurements of stellar age, stellar mass, and dust content. These discrepancies suggest that LAEs might have more complex, non-uniform evolutionary pathways, but stacked analyses can be challenging to interpret. Without understanding the shape of individual LAE star formation histories, it is difficult to understand the diversity in stellar mass assembly scenarios that can produce LAEs with these properties.


Following those studies, several groups made efforts to infer information about the rough shape of LAE SFHs. \citet{rosani2020bright} compared the instantaneous and average star formation rates of LAEs to show that some LAEs may be experiencing a ``rejuvenation event'' in which star formation recently restarted after the majority of their stellar mass was created. However, this study was unable to quantify the time, duration, amplitude, or frequency of the past star formation episode(s). The most recent studies of LAE star formation histories built on these analyses by testing SFHs of several predetermined functional forms, but distinguishing between different SFH models proved challenging \citep{iani2024midis, ceban2024star}. Due to the inevitable limitations of parametric SFH reconstruction techniques, these previous works have not been able to make definitive conclusions about the temporal processes and diversity of LAE stellar mass assembly.


To bring us closer to understanding the full picture of the physical nature of LAEs, it is now possible to use SEDs to uncover the detailed non-parametric histories of star formation and quenching throughout the lifetime of individual LAEs (and LBGs) without introducing significant bias.
The Dense Basis approach \citep{db1-iyer2017, db2-iyer2019} is a non-parametric Gaussian processes-based method for SED fitting and star formation history reconstruction \citep[see also][]{MOPED, MOPED_SFH, VESPA_SFH, prospector_leja2017deriving, prospector_johnson2021stellar}. 
The methodological improvements introduced in Dense Basis enable the reconstruction of realistic star formation histories with higher time resolution and a flexible number of peaks. This makes Dense Basis ideally suited for measuring the frequency and \edits{amplitude} of star formation episodes in LAEs and LBGs in order to \edits{better} understand if LAEs are uniquely experiencing their first major burst of star formation at the time in which we observe them.  

The One-hundred-deg$^2$ DECam Imaging in Narrowbands (ODIN) is a NOIRLab survey program designed to discover \edits{large} samples of LAEs at Cosmic Noon using narrowband imaging \edits{over extremely wide fields} \citep{odin_survey}. In order to create these LAE samples, ODIN utilizes three custom-made narrowband filters built for the Dark Energy Camera (DECam) on the V\'ictor M. Blanco 4m telescope at the Cerro Tololo Inter-American Observatory (CTIO) in Chile. These narrowband filters, \textit{N419}, \textit{N501}, and \textit{N673}, correspond to \edits{LAE} redshifts 2.4, 3.1, and 4.5
. Using these samples, we are able to identify protocluster\edits{s, trace cosmic filaments, and follow the evolution of} large scale structure during Cosmic Noon \citep[e.g.,][]{ramakrishnan2023odin, ramakrishnan2024odin_protoclusters_and_filaments, ramakrishnan2024odin_clustering_of_protoclusters, im2024testing}. With ODIN's final dataset, we anticipate a sample of $>$100,000 LAEs covering an area of $\sim$100 deg$^2$. 
Due to the vast photometric coverage in COSMOS from surveys like UVCANDELS \citep{candels1, candels2, UVCANDELS}, ODIN's LAE sample opens up many opportunities for synergistic research. 




In this work, we cross-match ODIN's LAE samples with the UVCANDELS photometric catalog in COSMOS to create sub-samples of LAEs with rest-UV-through-NIR photometry. With these data, we implement the Dense Basis approach to SED fitting and star formation history reconstruction 
and present the first analysis that can accurately characterize the diversity of LAE stellar mass assembly.
In Section \ref{sec:data} we introduce ODIN's LAE samples and the UVCANDELS photometric catalog. In Section \ref{subsec:SED_fitting} we discuss the Dense Basis methodology and introduce the SED fitting parameters and priors used in this analysis. In Section \ref{subsec:gold_samples} we discuss the generation of LAE samples and control LBG samples, and in Section \ref{subsec:measuring_peak_SF} we discuss star formation history analysis techniques. In Section \ref{sec:results} we introduce three sub-classes of LAE star formation histories, discuss the implications for LAE stellar mass assembly scenarios and progenitor histories, compare results to the control LBG samples, and re-evaluate \newedits{frequently made generalizations} about star formation in LAEs. The main conclusions of this work are summarized in Section \ref{sec:conc}. \edits{Throughout this work, we assume $\Lambda$CDM cosmology with $h$ = 0.7 and $\Omega_m$ = 0.27.}

\section{Data}\label{sec:data}


\subsection{ODIN LAE Samples}\label{subsec:ODIN_LAEs}

\citet{odin_lae_sel} introduced an improved technique for estimating Ly$\alpha$ emission line strength, the \textit{hybrid-weighted double-broadband continuum estimation} technique, as well as an innovative method for low redshift emission line galaxy interloper rejection. In short, LAEs were selected by looking for significant narrowband excess compared to the continuum estimated by two nearby broadband filters. When this narrowband excess corresponded to an estimated \edits{rest-frame Ly$\alpha$ EW} of 20 {\AA} or higher, that object was classified as an LAE\null. Utilizing a combination of narrowband excess and galaxy colors, several objects consistent with low-$z$ interlopers such as [\ion{O}{2}] emitters and Green Pea galaxies were eliminated from the LAE sample. With this new methodology, \citet{odin_lae_sel} introduced ODIN's inaugural sample of 6032, 5691, and 4066 LAEs at $z$ = 2.4, 3.1, and 4.5 \edits{, respectively,} in the extended COSMOS field ($\sim$9\,deg$^2$). These constitute some of the largest redshift-specific samples of narrowband-selected LAEs to date \edits{(comparable to those introduced by \citet{kikuta2023silverrush}). With ODIN's} samples, \citet{odin_lae_sel} presented scaled median stacked SEDs, revealing the overall success of ODIN's selection methods. Due to the vast array of rest-ultraviolet-through-near-infrared photometry available in the COSMOS field, we can search for photometric counterparts of ODIN's LAEs and use multi-wavelength photometry to fit model SEDs to our data.

\subsection{(UV)CANDELS Photometric Catalog}\label{subsec:CANDELS}

CANDELS \citep{candels1, candels2} is a photometric imaging survey aimed \edits{at probing} the galaxy population during Cosmic Noon (and Dawn) using the \textit{Hubble Space Telescope}'s near-infrared Wide Field Camera 3 (WFC3) and optical Advanced Camera for Surveys (ACS). Complementary to CANDELS, the UVCANDELS Survey\footnote{UVCANDELS data is available on the Mikulski Archive for Space Telescopes (MAST) with doi:\dataset[10.17909/8s31-f778]{\doi{10.17909/8s31-f778}} \citep{https://doi.org/10.17909/8s31-f778}.} \citep{UVCANDELS} introduced UV coverage to \edits{four} of the CANDELS fields with the WFC3/$F275W$ (with a depth of AB = 27) and ACS/$F435W$ (with a depth of AB $\geq$ 28.0) filters. Together with photometric data from CFHT \citep{cfht}, Subaru \citep{subaru1, subaru2}, \edits{Ultra}VISTA \citep{ultravista}, IRAC \citep{irac_seds, s-cosmos}, and NEWFIRM \citep{newfirm}, we have access to measurements from 46 filters (plus ODIN's three narrowbands). With these data, UVCANDELS has made photometric redshift measurements for each of their sources (see \citet{uvcandels_photozs} for details on photo-$z$ measurement techniques).

\section{Methodology}\label{sec:methods}

\subsection{Star Formation History Reconstruction}\label{subsec:SED_fitting}

A spectral energy distribution is a powerful tool for uncovering the history of star formation and quenching throughout the lifetime of a galaxy. In this work, we implement the Dense Basis non-parametric Gaussian process-based SFH reconstruction method of \citet{db1-iyer2017} and \citet{db2-iyer2019}. \edits{Parametric star formation histories typically allow for only a single peak of SFR, and for an actively star-forming galaxy like an LAE, this starburst is then highly likely to occur at the time of observation. For this reason, parametric models introduce significant biases in their solutions \citep[see][]{simha2014parametrising, carnall2019measure, dirichlet2}. In contrast, Dense Basis allows us to study star formation histories in much greater detail and with higher accuracy, including the possibility of multiple episodes of star formation. The Dense Basis method has been shown to yield non-parametric star formation history predictions that are robust enough to produce results consistent with the true properties of simulated galaxies, even in the presence of photometric noise (see Figure 6 in \citealt{db2-iyer2019}).} For example, Dense Basis is able to consistently measure the number of peaks in a galaxy's star formation history, even with a varying number of parameters (see Figure 7 in \citealt{db2-iyer2019}). Additionally, Dense Basis is able to reconstruct star formation histories with a scatter of 0.2 dex out to a lookback time of $\sim$5 Gyr \citep{db2-iyer2019}. This is more than sufficient for our sample, which goes out to a maximum lookback time of $\sim$2.8 Gyrs. These advantages highlight the suitability of Dense Basis for addressing the key questions of this work without significant systematic limitations.

In order to generically go from an SED to a star formation history, we first fit SED datapoints to Flexible Stellar Population Synthesis
\citep[FSPS;][]{FSPS1, FSPS2} templates with well-understood \edits{properties}. \edits{We then assume a stellar metallicity, a stellar initial mass function, a dust attenuation model, and an absorption law for the effects of the intergalactic medium. Each of these \edits{assumptions} helps turn a generic stellar population into a realistic galaxy model. We then compare the resultant model SED to observations and use spectral template features such as Ly$\alpha$ for recent star formation, emission lines like H$\beta$ and [\ion{O}{3}] for ISM conditions, long wavelengths for past star formation, and IR wavelengths for dust re-emission. These features allow us to infer the rate of star formation vs. time. SFHs enable us to uncover the physical nature of galaxies and better understand their life-stories and global properties. For a more detailed description of the Dense Basis SED fitting and SFH reconstruction methodology, please refer to Section 2 of \citet{db2-iyer2019}.} 

For our SFH reconstruction via SED fitting, we adopt the fiducial priors from the observed properties of LAEs \citep[e.g.,][]{gawiser2007lyalpha, guaita2011lyalpha, vargas2014stack, acquaviva2012curious} as follows. We assume the log stellar mass \edits{in units of solar masses $\log M_*/M_\odot$} is uniform from 7 to 12, the log of the instantaneous star formation rate $SFR$ is uniform from $-$1.0 to 2.0, log(metallicity/$Z_\odot$) is uniform from $-$1.5 to 0.2, and the prior for the lookback times ${t_x}$ is Dirichlet with alpha = 3.0 \citep{prospector_leja2017deriving, dirichlet2}. We also assume Chabrier IMF \citep{chabrier2003galactic} and a Calzetti dust model \citep{calzetti2000dust} with a flat dust extinction $A_V$ prior from 0.0 to 2.0. Lastly, we assume a flat redshift distribution, confined by the full-width half-max (FWHM) of ODIN's narrowband filter transmission curves \citep{odin_survey}. \edits{We note that these FWHM measurements are in good agreement with the redshift distributions of the DESI-confirmed ODIN $N419$- and $N501$-selected LAEs (see Figure 3 of \citet{white2024clustering}).} We find that our posterior distributions are much narrower than the top-hat priors and do not push up against the bounds of the priors, indicating that \edits{our priors are reasonable and} do not introduce significant bias in our measurements. \edits{Further, we find that changing the number of ${t_x}$ parameters used in the priors would not produce any notable difference the SFH archetype classification or number of major SF bursts in the samples.}

To carry out our SED fitting, we begin with all CANDELS and ODIN filters outlined in Table \ref{tab:filters}. We chose to exclude IRAC channels 3 and 4 since their large PSF and photometric errors often result in strong disagreement with other photometric bands, artificially driving up SED-fit reduced $\chi^2$ values. We also \edits{remove} any narrow- and intermediate band filters that contain the Ly$\alpha$ emission line for a given object. This may initially seem counter-intuitive, however, we should not fully trust the Ly$\alpha$ produced by stellar population synthesis models since it does not capture the complexity of Ly$\alpha$ radiative transfer. \edits{Therefore, i}n addition to the relevant ODIN narrowband, for the $z$ = 2.4, 3.1, and 4.5 samples we eliminate the $IA427$, $IA505$, and $IA679$ intermediate band filters, respectively. \edits{Finally, we introduce observed photometric error floors, which have been found to successfully represent a combination of systematic errors in photometric calibration and SED template incompleteness \citep{iyer2019reconstructing}.}
\begin{equation}\label{eq:error}
    obs_{err} = \sqrt{(f_{err}^2 + (U\times f)^2)},
\end{equation}
where $f_{err}$ is the measured flux error for a particular filter, $U$ is a fractional zeropoint uncertainty, $f$ is the measured flux for a particular filter, and $obs_{err}$ is the new observed error used in the SED fitting. For space-based observations we use $U$ = 0.03 \edits{and} for ground-based observations and IRAC data we use $U$ = 0.10\edits{, as adapted from \citep{iyer2019reconstructing}}. This allows us to give more weight to space-based photometry, for which we have \edits{less variation in observing conditions and} greater confidence. 

\subsection{LAE \& Control LBG Samples}\label{subsec:gold_samples}

In order to ensure that our samples are as pure as possible, we create LAE and control LBG samples utilizing SED-fit reduced $\chi^2$ values and\edits{, in the case of the LBGs,} UVCANDELS photo-$z$ data. 

For our LAE sample, we begin by taking each ODIN LAE in COSMOS that has UVCANDELS photometry. 
We then only retain objects with Dense Basis SED reduced $\chi^2$ values $\leq$ 2.6. We determined this cutoff by matching the maximum SED-fit reduced $\chi^2$ for an object with spec-$z$ confirmation. We find that this corresponds well to the transition between objects with and without strong UVCANDELS photo-$z$ agreement. 
We note that the \edits{majority of the} discarded \edits{objects} are \edits{likely} still LAEs since our narrowband photometry, which can be crucial for constraining redshift, is not included in the SED fitting or photo-$z$ calculation. However, we choose to leave them out of our sample to ensure sample purity, thereby reducing the risk of misleading conclusions from outliers. \edits{This reduced $\chi^2$ cut also limits any potential contamination from active galactic nuclei, which tend to have very poor SED fits.} In order to make sure this method does not introduce any strong systematic biases into our data, we examine the \edits{magnitude} distributions \edits{in} $ugrizy$ \edits{and} narrowband \edits{filters} for the LAEs \edits{with $\chi^2 \leq 2.6$ compared to those with $\chi^2 > 2.6$.}
We find no indications of systematic bias. We also examine the EW and Ly$\alpha$ line flux distributions of the $\chi^2$-limited sample and find that the majority of the eliminated objects are on the lower EW/Ly$\alpha$ line flux end of the samples\edits{, though many low EW/Ly$\alpha$ line flux LAEs still remain}. The median EWs of our LAE samples thus only change modestly, from (52, 46, 62)\,{\AA} to (58, 54, 62)\,{\AA} for $z$ = 2.4, 3.1, and 4.5, respectively. Similarly, the median Ly$\alpha$ line fluxes in units of 
$10^{-17}$~erg/s/cm$^2$ change from (3.2, 2.7, 1.6) to (3.3, 2.7, 1.6), respectively.
This indicates that we are protected against low-$z$ interlopers, which often have lower EWs. 
The $\chi^2$ cut removes 11, 11, and 7 objects, leaving the final samples for $z$ = 2.4, 3.1, and 4.5, respectively.



Companion to our LAE samples, we create control samples of generic star forming galaxies with comparable redshift by selecting galaxies primarily based on their Lyman Break, \edits{i.e.} Lyman Break Galaxies (LBGs). To create the control LBG samples, we follow a more stringent procedure \edits{compared} to that of the LAE samples since ODIN's LAE samples \edits{have already undergone} a rigorous selection process utilizing ODIN narrowband photometry \citep[see][]{odin_lae_sel}. To accomplish this, we utilize UVCANDELS photo-$z$ measurements \citep{uvcandels_photozs}. In \edits{general}, the enhanced UV coverage helps to better constrain photo-$z$ measurements and increase confidence in our sample purity, especially for the $z$ = 2.4 sample for which $F275W$ is particularly helpful for picking up the Lyman break ($\sim$310\,nm). 

For each object in COSMOS that has UVCANDELS photometry, we first integrate the UVCANDELS redshift probability density function (PDF) over a redshift range twice as broad as the FWHM of each of ODIN's filter transmission curves with the same \edits{median redshift}. We chose this range since it is much more difficult to constrain LBG redshift\edits{s} without narrowband imaging. This also helps us to retain a reasonable sample size for comparison without sacrificing sample purity. We then divide \edits{the integral over this range} by the total integrated PDF to get the probability that a given object falls into that redshift range. From this sample, we select all objects that have a $\geq$ 68\% probability of falling into this range. \edits{These steps ensure that the LBG sample has a comparable redshift range to ODIN's LAE sample.} We retain objects with SED reduced $\chi^2$ values $\leq$ 2.6 \edits{(for uniformity with the LAEs)} and remove a small fraction of objects that are already classified as LAEs (roughly 3\%). Lastly, we set a maximum stellar mass limit such that the median stellar mass of each control LBG sample matches the corresponding LAE sample. \edits{Although LAEs are typically younger and lower-mass than LBGs, we chose to apply a mass limit to the LBG sample in order to best compare fundamental galaxy properties and limit any biases introduced by differences in either stellar mass or the resulting total SED signal to noise ratio for otherwise similar galaxies.} After these steps, we find 
27, 160, and 74
LBGs in the control samples for $z$ = 2.4, 3.1, and 4.5, respectively. We also retain the higher-mass LBG sample at each redshift to be used for assessing overall LBG sample characteristics (see Figure \ref{fig:full_tot_stellar_mass_comp_LAE_LBG}). 



\subsection{Measuring The Largest Burst(s) of Star Formation}\label{subsec:measuring_peak_SF}

In order to determine the time of a galaxy's most prominent period of star formation, we search for the SFH peak corresponding to the galaxy's maximum star formation rate. If this peak coincides with the time of observation (with a resolution of 200\,Myr \edits{due to the near-UV SFR timescale found by \citealt{broussard2019star}}), then we conclude that a galaxy is currently experiencing its most significant burst of star formation thus far. To validate this method, we also determine the largest cumulative stellar mass generated within $\pm$100\,Myr of an SFR peak. We find that this would make no significant difference in results except in discriminating against bursts that have not ramped down by the time of observation in cases where there is significant stellar mass assembly in the past. 


We also determine the fraction of cumulative stellar mass created at the time of observation, with a resolution of 200\,Myr. We accomplish this by integrating the star formation rate as a function of time to compute a cumulative stellar mass formed (not accounting for stellar death) over the last 200\,Myr, and then divide that by the total stellar mass \edits{accumulated} over the lifetime of each galaxy.  

\section{Results}\label{sec:results}


\edits{As a quick reality check, we compare the median values for $\log(SFR/(M_\odot/$yr$))$, $\log(M_*/M_\odot)$, and $A_V$ for each sample at each redshift. For $\log(SFR/(M_\odot/$yr$))$, we find median values of 0.72$\pm 0.3$, 0.77$\pm 0.2$, and 1.2$\pm 0.3$ for $z$ = 2.4, 3.1, and 4.5, respectively. For $\log(M_*/M_\odot)$, we find median values of 8.6$\pm 0.6$, 8.8$\pm 0.2$, and 9.3$\pm 0.3$, respectively. For $A_V$, we find median values of 0.5$\pm $0.2, 0.5$\pm 0.1$, and 0.5$\pm 0.1$, respectively.} These values are comparable to those presented in previous works \citep[e.g.,][]{gawiser2007lyalpha, guaita2011lyalpha, acquaviva2012curious, vargas2014stack, kusakabe2018stellar}.

\subsection{LAE Star Formation Histories}\label{subsec:LAE_SFHs}

\begin{figure*}
\begin{center}
\includegraphics[width=1\textwidth]{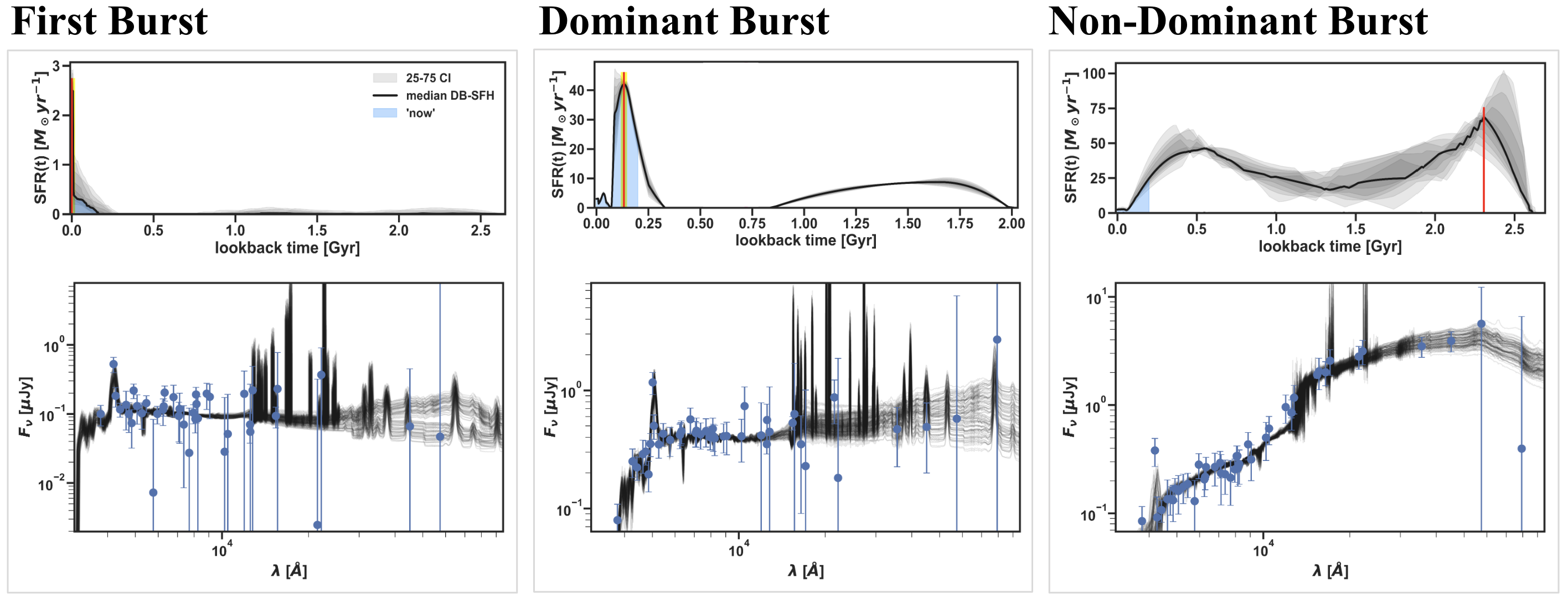}
\caption{Examples of three classes of observed LAE star formation histories \edits{and their SED fits}: \textit{first burst}, \textit{dominant burst}, and \textit{non-dominant burst}. Each LAE used in this figure has a confirmed spectroscopic redshift. In each star formation history (top), the x-axis represents the lookback time in Gyrs and the y-axis represents the star formation rate as a function of time in solar masses per year. The black solid line represents the median Dense Basis star formation histories. The gray shaded region represents the 25-75\% confidence interval for the star formation history. The blue shaded region represents the last 200\,Myr, \edits{i.e.} the age range of stars able to dominate the rest-UV light at the time of observation. The red solid vertical line represents the time at which the galaxy reached its maximum star formation rate. In cases where the maximum star formation rate occurs `now' \edits{(in the last 200\,Myr)}, it is highlighted in yellow. \edits{In each SED fit (bottom), the x-axis represents the wavelength $\lambda$ in Angstroms and the y-axis represents the flux density $F_\nu$ in microjansky. The blue points represent the photometric data, and the black lines represent the spectral template models used to determine the median SFH and confidence intervals in the above panels.}}
\label{fig:SFH_types}
\end{center}
\end{figure*}

\edits{For this paper, we consider three different possible LAE star formation history archetypes (see Figure \ref{fig:SFH_types}). The first is that associated with the \newedits{frequently assumed} paradigm for LAEs, i.e., galaxies undergoing their \textit{first burst} of star formation (SFR $\ge 1 M_{\odot}$~yr$^{-1}$) at the time of observation. Alternatively, a galaxy may be undergoing a \textit{dominant burst} of star formation; in these objects, significant (SFR $\ge 1 M_{\odot}$~yr$^{-1}$) bursts of star formation occurred in the past, but none of these bursts are as large as the present SFR\null. The third category is for \textit{non-dominant burst} LAEs; in these objects, the greatest SFR occurred in the past.}

\begin{figure*}
\begin{center}
\includegraphics[width=1\textwidth]{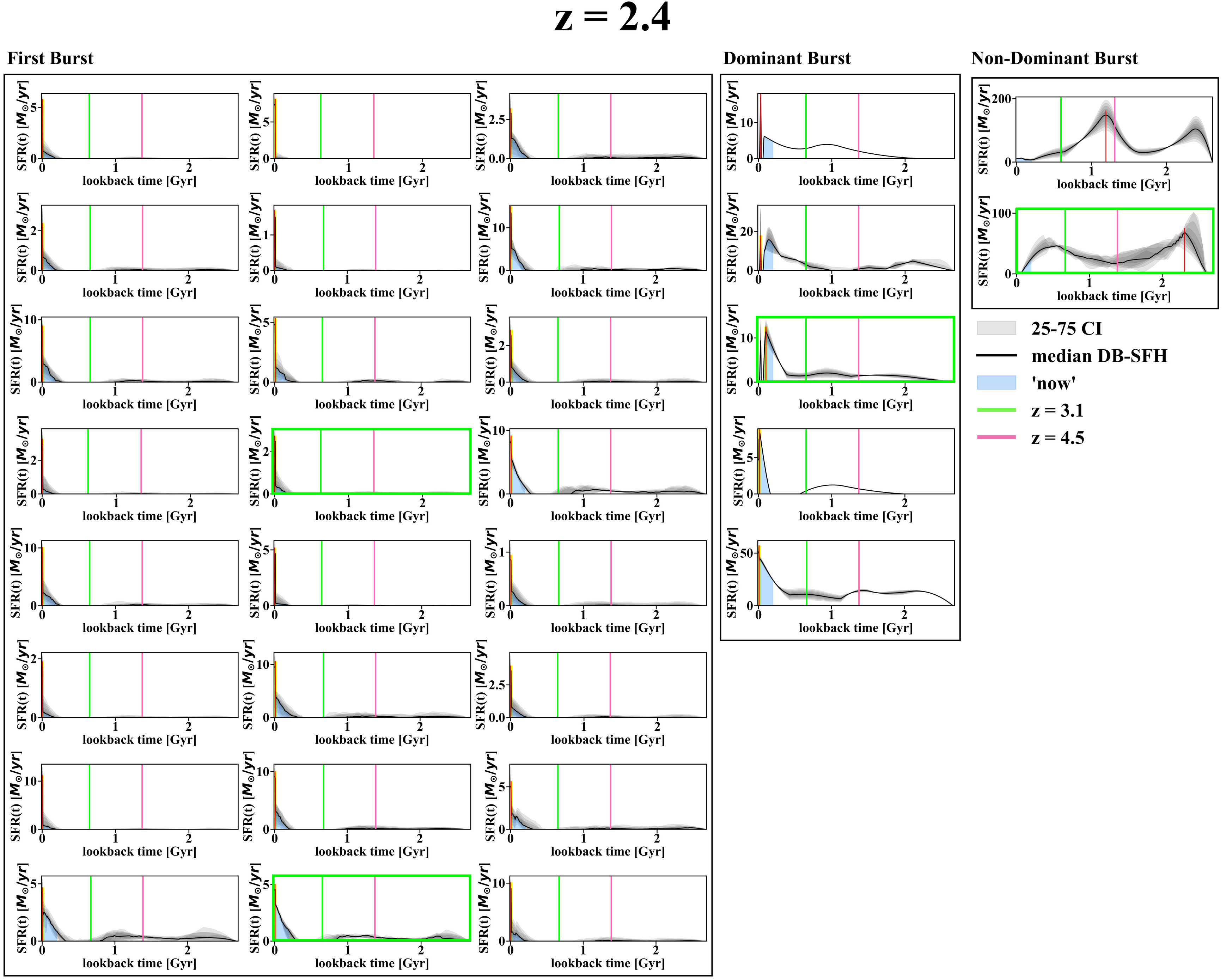}
\caption{Star formation histories of $z$ = 2.4 LAEs. The specifications in this figure match those of the star formation histories in Fig.~\ref{fig:SFH_types}. Plots outlined in green highlight LAEs with spectroscopic confirmations. The green and pink vertical lines (when present) represent the lookback times corresponding to $z$ = 3.1 and 4.5, respectively. In this sample there are 24 LAEs with \textit{first burst} SFHs, 5 LAEs with a \textit{dominant burst} SFH, and 2 LAEs with \textit{non-dominant burst} SFHs.} 
\label{fig:2.4_SFH}
\end{center}
\end{figure*}

\begin{figure*}
\begin{center}
\includegraphics[width=1\textwidth]{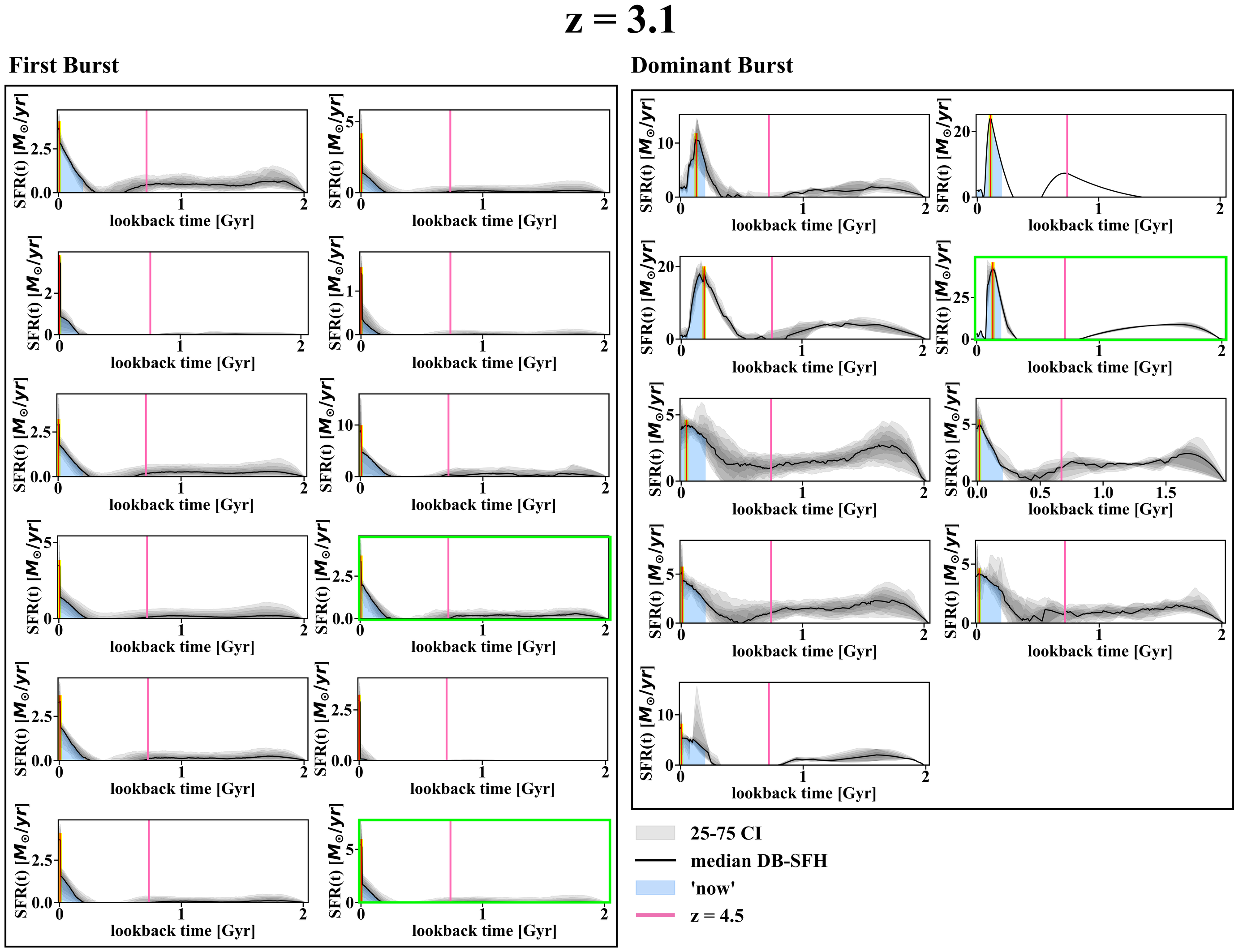}
\caption{Star formation histories of $z$ = 3.1 LAEs. The specifications in this figure match those of the star formation histories in Fig.~\ref{fig:SFH_types}. In this sample there are 12 LAEs with \textit{first burst} SFHs and 9 LAEs with \textit{dominant burst} SFH.} 
\label{fig:3.1_SFH}
\end{center}
\end{figure*}

\begin{figure*}
\begin{center}
\includegraphics[width=1\textwidth]{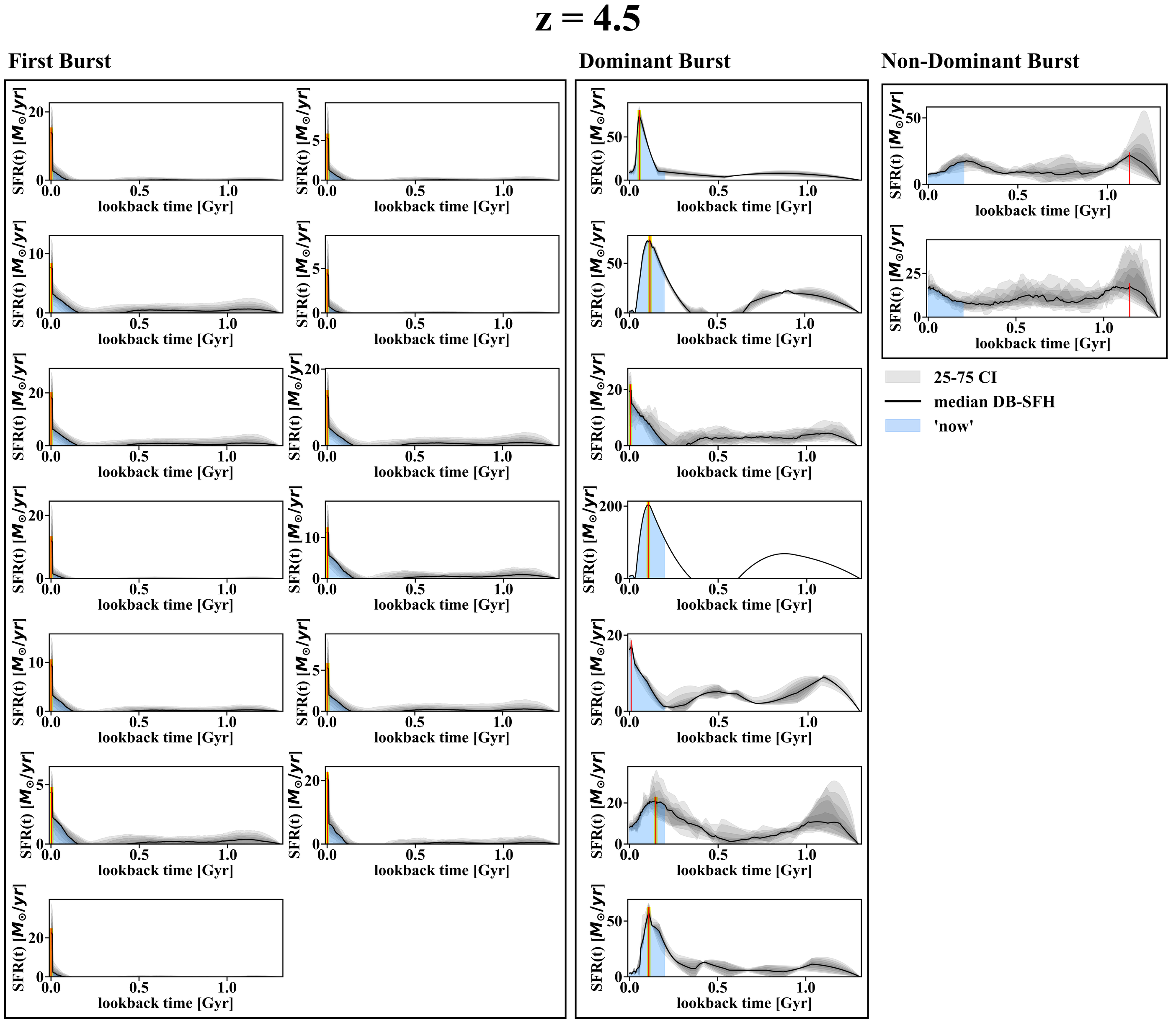}
\caption{Star formation histories of $z$ = 4.5 LAEs. The specifications in this figure match those of the star formation histories in Fig.~\ref{fig:SFH_types}. In this sample there are 13 LAEs with \textit{first burst} SFHs, 7 LAEs with \textit{dominant burst} SFH, and 2 LAEs with \textit{non-dominant burst} SFHs.} 
\label{fig:4.5_SFH}
\end{center}
\end{figure*}

At $z$ = 2.4, 3.1, and 4.5, respectively, we find that 77\%, 57\%, and 59\% of LAEs fall into the \textit{first burst} category; 16\%, 43\%, and 32\% fall into the \textit{dominant burst} category; and 6\%, 0\%, and 9\% fall into the \textit{non-dominant burst} category (see Figures \ref{fig:2.4_SFH}-\ref{fig:4.5_SFH} for individual star formation histories for each LAE sample). These results suggest that undergoing a first major burst of star formation \edits{at the time of observation} is quite common but is not a prerequisite for a galaxy to be classified as an LAE\null. We can, therefore, conclude that there must be multiple stellar mass \edits{assembly scenarios capable of exhibiting} strong Ly$\alpha$ emission at the time of observation. 



\edits{While \textit{dominant burst} LAEs are not entirely out of the realm of expected LAE SFHs, it would have been highly surprising if \textit{non-dominant} burst LAEs were common in the samples. This is because LAEs are selected for ongoing star formation without significant quenching of Ly$\alpha$ photons by dust. That being said, even the presence of the four observed \textit{non-dominant} bursts is worth trying to understand (despite the small number statistics of finding two such LAEs at $z=2.4$ and two more at $z=4.5$ prohibiting us from being able to make precise measurements about just how rare this type of LAEs is at each redshift). We note that the minimum time-of-observation SFR of our LAEs is not particularly high, roughly $1\,M_{\odot}$~yr$^{-1}$. For more massive galaxies like those seen as \textit{non-dominant} bursts at $z$ = 2.4, this does not require anything close to a peak of star formation. A lack of dust is harder to imagine in such galaxies. This implies that the regions of such galaxies from which Ly$\alpha$ photons are escaping have either had most of their dust destroyed and not yet re-formed in a new starburst, or that the anisotropic radiative transfer of those Ly$\alpha$ photons is fortuitous in allowing them to escape towards Earth.} 

\edits{Further, w}e find that the $z$ = 4.5 LAEs with \textit{non-dominant burst} star formation histories tend to have maximum bursts of star formation with similar amplitudes to LAEs with \textit{first burst} and \textit{dominant burst} star formation histories at the same redshift. \edits{In contrast}, the $z$ = 2.4 LAEs with \textit{non-dominant burst} star formation histories experienced bursts of star formation that are an order of magnitude higher in amplitude than those of \textit{first burst} and \textit{dominant burst} LAEs. We interpret \edits{these 2 LAEs} as possible evidence of several \edits{past} galaxy mergers, which \edits{have built the galaxies' current stellar mass} over the course of 2.8 billion years. 
That being said, there is still more investigation \edits{into the morphology of these sources that is} needed to draw more definitive conclusions about the likelihood of these scenarios. \edits{For this reason, we encourage the reader to interpret these LAEs with \textit{non-dominant burst} SFHs as interesting case studies that highlight the diversity in possible LAE stellar mass assembly scenarios rather than representations of all \textit{non-dominant burst} LAEs within the overall population.}



\begin{figure*}
\begin{center}
\includegraphics[width=1\textwidth]{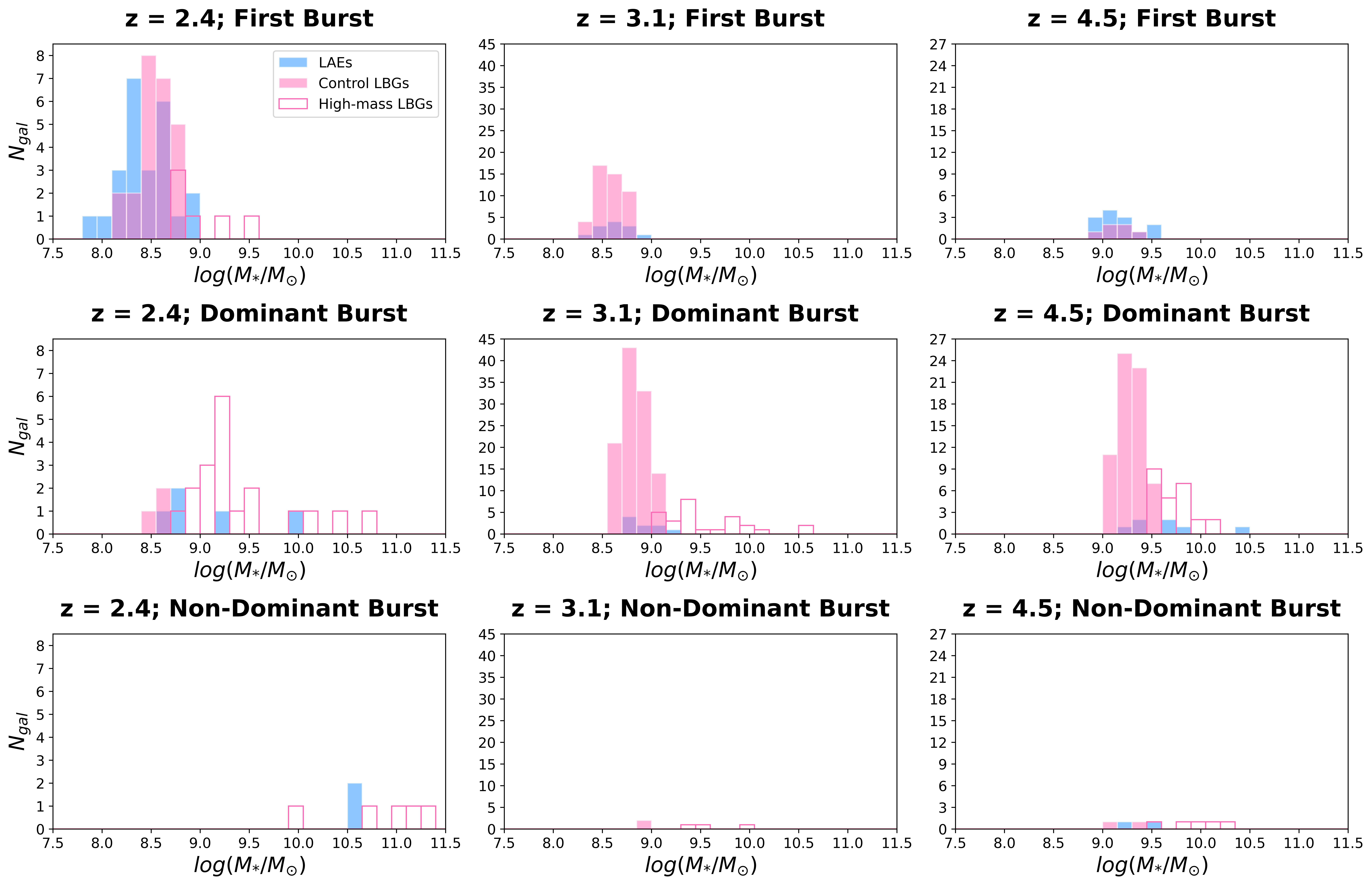}
\caption{
Total stellar mass histograms for three star formation history archetypes of LAEs and LBGs at each redshift. The columns represent $z$ = 2.4 (left), 3.1 (middle), and 4.5 (right). The rows represent the \textit{first burst} (upper), \textit{dominant burst} (middle), and \textit{non-dominant burst} (lower) archetypes. In each histogram, the x-axis represents the log of the total stellar mass $M_*$ in solar masses. The y-axis represents the number of galaxies $N_{gal}$. The blue histograms represent the LAEs and the pink histograms represent the LBGs, with the hollow pink histograms denoting random samples of 30 high stellar mass LBGs that were not included in the LBG control sample.}
\label{fig:full_tot_stellar_mass_comp_LAE_LBG}
\end{center}
\end{figure*}

In Figure \ref{fig:full_tot_stellar_mass_comp_LAE_LBG}, we present the total stellar mass $M_*$ distribution for LAEs compared to LBGs within each star formation history archetype. For further comparison, we \edits{also plot} random subsets of high stellar mass LBGs that were not included in the stellar mass-limited sample. 
The breakdown of objects in each subplot of Figure \ref{fig:full_tot_stellar_mass_comp_LAE_LBG} reveals that LAEs preferentially fall into the \textit{first burst} archetype while LBGs preferentially fall into the \textit{dominant burst} archetype, hinting at a characteristic difference in stellar mass assembly scenarios that are likely to produce a galaxy detected as an LAE vs. an LBG. We find that the total stellar mass accumulated, $M_*$, increases modestly as we move from \edits{\textit{first burst} to \textit{dominant burst}, to \textit{non-dominant burst}}, telling us that the functional behaviors of the star formation histories also generally correlate with cumulative stellar mass. 

\subsection{Formative Bursts of Star Formation}\label{subsec:formative_burst}

In this section we refer to all stellar mass assembly in the last 200\,Myr as ``the time of observation'' since the UV SFR is a weighted average \edits{over} the last 200\,Myr SFR. We find that 
94\%, 100\%, and 91\% of LAEs are in their maximum period of star formation since the Big Bang at the time of observation at $z$ = 2.4, 3.1, and 4.5, respectively, compared to 100\%, 99\%, and 97\% of the corresponding LBG samples. \edits{The differences are not significant compared to Poisson uncertainties in the numbers of galaxies with \textit{non-dominant bursts} at each epoch. }

\begin{figure*}
\begin{center}
\includegraphics[width=\textwidth]{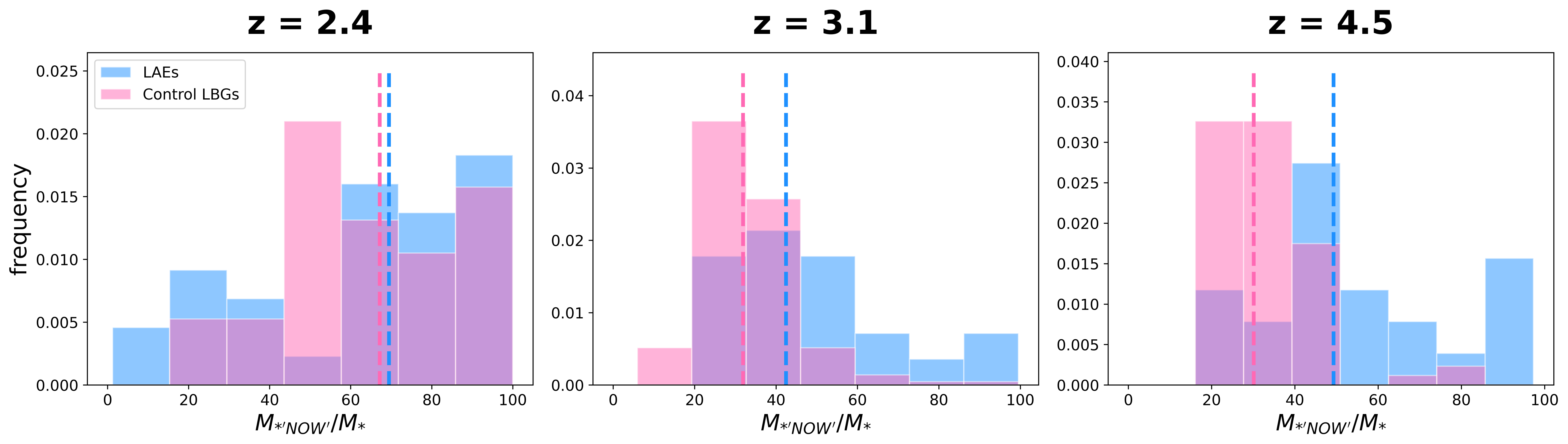}
\caption{Histograms showing the percentage of stellar mass created at the time of observation for LAEs compared to LBGs. The x-axis represents the percentage of stellar mass created at the time of observation (the last 200\,Myr) $M_{*\text{`NOW'}}/M_{*}$. The y-axis represents the \edits{\textit{normalized} frequency of each type of galaxy}. The blue histograms represent the LAE samples and the pink histograms represent the control LBG samples. The vertical dashed lines represent the corresponding median fraction of stellar mass created at the time of observation for each population. These plots demonstrate that the median fraction of stellar mass created at the time of observation is consistently larger for LAEs than LBGs, although the difference becomes insignificant at $z$ = 2.4.} 
\label{fig:formative_burst_comp}
\end{center}
\end{figure*}

\edits{We further investigate the stellar mass assembly by measuring the median fraction of stellar mass created during the time of observation $Med(M_{*\text{`NOW'}}/M_{*})$ for LAEs compared to LBGs. We find that} $Med(M_{*\text{`NOW'}}/M_{*})$ is 
1.03, 1.33, and 1.64 times higher in LAEs than LBGs at $z$ = 2.4, 3.1, and 4.5, respectively (see Figure \ref{fig:formative_burst_comp}). \edits{To test the statistical significance of the difference in the distribution of $M_{*\text{`NOW'}}/M_{*}$ measurements for LAEs vs. LBGs, we perform a Kolmogorov-Smirnov test \citep[e.g.,][]{KS_test}, which returns p-values of 0.80, 1.8E-04, and 1.7E-05, respectively. We conclude that the null hypothesis of being drawn from the same underlying distribution is rejected with very high confidence for the $z$ = 3.1 and 4.5 samples.} The null hypothesis of being drawn from the same underlying distribution is not rejected for the $z$ = 2.4 samples. This suggests that LBGs at $z$ = 2.4 are more like LAEs at $z$ = 2.4 than the higher-$z$ LBGs, however it may also be a consequence of the smaller LBG sample size, making it more difficult to distinguish any noticeable differences between the LAEs and LBGs. Overall, these results suggest that LAEs typically assemble more \edits{fractional} stellar mass than LBGs in the 200\,Myr prior to observation, although the quantitative difference is not dramatic.

Examining Figure \ref{fig:formative_burst_comp} in more detail, we note that very few objects' SFHs are consistent with forming 100\% of their stellar mass within the last 200\,Myr.  Even though \textit{first burst} LAEs represent a majority, the SFHs typically imply low-level star formation at earlier times, with SFRs too low for us to classify as ``bursts.''  This is not unexpected given hierarchical structure formation; even in a region of the universe that is just forming an LAE at the time of observation, it would be quite surprising if no low-mass galaxies with existing stars were present. Additionally, we see a higher fraction of recent stellar mass in the lower-redshift samples (see Figure \ref{fig:formative_burst_comp}). We hypothesize that this is because the mass range of our LAE (and hence LBG control) samples is evolving downwards (due to luminosity distance squared) faster than galaxy properties are evolving. 

The broader characterization of LAE star formation discussed in this section leads us to the conclusion that most LAEs are best described as experiencing a \textit{formative burst} of star formation, ie. a burst that played a major role in the formation of the galaxy being observed. We designate this significant subset of LAEs ($\sim$95) by combining the galaxies with \textit{first burst} and \textit{dominant burst} SFH archetypes. The results discussed in this section also elucidate clear differences between the LAE and LBG samples, however they still do not provide a full picture of the fundamental differences driving the detection of a galaxy as an LAE vs. an LBG\null. Further investigation into radiative transfer mechanisms and ISM geometry including the opening angle for \edits{Ly$\alpha$} emission will be necessary to more fully develop this picture.  

We also note that these results are not changed substantially if we reconsider the objects removed by our $\chi^2$ cut. The overall distribution between archetypes only changes modestly, with about 10\% more of the LAEs in the dominant burst category, but still $\sim$95\% of LAEs experiencing formative bursts. Spectroscopic follow-up is needed to determine if the LAEs with poor $\chi^2$ are truly interlopers or lower-EW LAEs; at this point, because the SED fits are poor, it remains safer to exclude these objects.

\subsection{Is the LAE Phenomenon Long- Lived?}\label{subsec:long-lived}

\edits{While we find that the vast majority of LAEs are experiencing a \textit{formative} burst of star formation, the prevalence of past star formation in our LAE SFHs motivates the question of whether or not an LAE had sufficient past star formation to plausibly have been an LAE at earlier times as well.} Prior to this work, studies have suggested that LAEs could have multiple Ly$\alpha$ emitting phases \citep[see][]{hathi2016vimos}, however previous techniques have been limited in their ability to test this hypothesis. We \edits{investigate this possibility} by comparing the star formation rates at earlier times to those of our LAEs at the time of observation. When the star formation rate of an LAE at any given lookback time is comparable to the minimum time-of-observation star formation rate of our LAE samples, $\sim$1\,$M_\odot$~yr$^{-1}$, we consider this object to have \edits{\textit{possibly}} been an LAE at an earlier time as well. We note that this star formation rate threshold should be interpreted as a necessary but far from sufficient condition for these objects to be classified as an LAE at earlier times. This is because the star formation rate is not the only factor that contributes to a galaxy being classified as an LAE since it does not incorporate the nuances of radiative transfer and the orientation of the line-of-sight to Earth.

We find that only 23\% (7) of the $z$ = 2.4 LAEs had star formation rates exceeding 1\,$M_\odot$~yr$^{-1}$ at some earlier time, implying that they could plausibly have been observed as LAEs at that earlier epoch.  
Applying the same SFR threshold, 6 of those $z$ = 2.4 LAEs could have plausibly been LAEs at $z$ = 3.1 and 5 at $z$ = 4.5. For the $z$ = 3.1 LAE sample, 43\% (9) had star formation rates exceeding 1\,$M_\odot$~yr$^{-1}$ at some earlier time; 5 of those LAEs could have plausibly been LAEs at $z$ = 4.5. And for the $z$ = 4.5 LAE sample, 41\% (9) had star formation rates exceeding 1\,$M_\odot$~yr$^{-1}$ at some earlier time. Finding similar numbers for specific epochs vs. any earlier epoch matches the visual impression from Figures \ref{fig:2.4_SFH}-\ref{fig:4.5_SFH} that a subset of each observed LAE sample has SFHs with high SFR over a long time period. Together, these results suggest that ``LAE'' should be viewed as a classification that is both temporary and of varying duration, and that a galaxy may possibly be an LAE (or have LAE progenitors) at multiple points during its lifetime.

\section{Conclusions}\label{sec:conc}

The Dense Basis method introduced state-of-the-art advancements to star formation history reconstruction, enabling the flexibility to model smooth star formation histories with several periods of star formation and quenching \citep{db1-iyer2017, db2-iyer2019}. We applied this approach to LAEs discovered in the COSMOS field by ODIN, a narrow-band survey program designed to discover $z=2.4, 3.1$, and 4.5 Ly$\alpha$ emission over extremely wide fields.

In this work, we paired data from the ODIN narrowband LAE survey with rest-UV-through-NIR photometry and photometric redshifts from UVCANDELS to produce samples of LAEs and control samples of LBGs at $z$ = 2.4, 3.1, and 4.5. With the Dense Basis methodology and galaxy samples, we reconstructed star formation histories and explored the stellar mass assembly of LAEs in detail. This allowed us to test the \newedits{frequently adopted characterization of} LAEs \newedits{as} young galaxies uniquely experiencing their first significant period of star formation at the time of observation and better understand how LAE stellar mass assembly compares to other star forming galaxies. The main conclusions of this work are summarized below. 
\begin{enumerate} [leftmargin=1.3\parindent] 

    \item LAE star formation histories can be characterized by three archetypes: \textit{first burst} at the time of observation, \textit{dominant burst} showing the highest SFR at the time of observation but with at least one significant star formation episode in the past, and \textit{non-dominant burst} describing the case where the highest SFR occurred in the past. While the majority of LAEs at all three redshifts have \textit{first burst} SFHs, a significant fraction at each redshift has \textit{dominant burst} SFHs, and \textit{non-dominant burst} SFHs are quite rare. These results show that a first major burst of star formation is not a prerequisite for a galaxy to be discovered as an LAE\null. Multiple stellar mass assembly scenarios are consistent with a galaxy \edits{exhibiting} strong Ly$\alpha$ emission at the time of observation. 


    \item \edits{Within LAE and LBG ``control'' samples with the same median stellar mass,} the majority of LAE SFHs exhibit  the \textit{first burst} archetype at all three redshifts, as do the majority of control LBGs at $z$ = 2.4, while the majority of control LBGs at $z$ = 3.1 and $z$ = 4.5 (and the majority of massive LBGs at all three redshifts) have SFHs exhibiting the \textit{dominant burst} archetype. This demonstrates that LAE stellar mass assembly differs from other star-forming galaxies.

    \item The \edits{fraction} of stellar mass created by LAEs at the time of observation (last 200\,Myr) is on average 1.3 times that of LBGs at comparable redshifts \edits{and stellar masses}, affirming that LAEs generally have less evolved stellar populations than other star-forming galaxies. 

    \item By examining the characteristics of LAE stellar mass assembly outlined above, we conclude that the vast majority ($\sim$95) of LAEs are best described as  
    experiencing a \textit{formative} burst of star formation, i.e., a burst that played a major role in the formation of the galaxy being observed. 

    \item On average 34\% of LAEs have star formation rates at some earlier period \edits{that are} comparable to the minimum time-of-observation star formation rate of our LAE samples. These results suggest that ``LAE'' should be viewed as a classification that is both temporary and of varying duration, and that a galaxy may be an LAE (or have LAE progenitors) at multiple points during its lifetime.

\end{enumerate}

Overall, our results suggest that the \newedits{historical characterization} of LAE stellar mass assembly is incomplete and that there are several evolutionary paths that \edits{lead to the formation of galaxies with strong} observed Ly$\alpha$ emission. With this \edits{work}, we begin to understand what makes LAEs unique probes of the high redshift universe, opening up new avenues to observe LAE stellar mass assembly, progenitor history, and radiative transfer in the post-reionization universe. 


\newpage

\section{Acknowledgments}


This work utilizes observations at Cerro Tololo Inter-American Observatory, NSF’s NOIRLab (Prop. ID 2020B-0201; PI: KSL), which is managed by the Association of Universities for Research in Astronomy under a cooperative agreement with the NSF.

This material is based upon work supported by the NSF Graduate Research Fellowship Program under Grant No. DGE-2233066 to NF. NF and EG acknowledge support from NSF grant AST-2206222 and NASA Astrophysics Data Analysis Program grant 80NSSC22K0487. NF would like to thank the LSST-DA Data Science Fellowship Program, which is funded by LSST Discovery Alliance, NSF Cybertraining Grant 1829740, the Brinson Foundation, and the Moore Foundation; her participation in the program has benefited this work greatly.  
EG acknowledges the support of an IBM Einstein fellowship for his sabbatical at IAS during the completion of this manuscript. 
\edits{KSL and VR acknowledge financial support from the NSF under Grant Nos. AST-2206705, AST-2408359, and AST-2206222, and from the Ross-Lynn Purdue Research Foundations.}
\edits{RC and CG acknowledge support from NSF grant AST-2408358.} The Institute for Gravitation and the Cosmos is supported by the Eberly College of Science and the Office of the Senior Vice President for Research at the Pennsylvania State University.
LG 
thanks support from FONDECYT regular proyecto No. 1230591.
LG also gratefully acknowledges financial support from ANID - MILENIO - NCN2024\_112 and from the ANID BASAL project FB210003. HSH acknowledges the support of the National Research Foundation of Korea (NRF) grant funded by the Korea government (MSIT), NRF-2021R1A2C1094577, Samsung Electronic Co., Ltd. (Project Number IO220811-01945-01), and Hyunsong Educational \& Cultural Foundation.
YY is supported by the Basic Science Research Program through the National Research Foundation of Korea funded by the Ministry of Science, ICT \& Future Planning (2019R1A2C4069803).
\edits{JL is supported by the National Research Foundation of Korea (NRF-2021R1C1C2011626).} HS acknowledges the support of the National Research Foundation of Korea grant, No. 2022R1A4A3031306, funded by the Korean government (MSIT). \edits{XW is supported by the National Natural Science Foundation of China (grant 12373009), the CAS Project for Young Scientists in Basic Research Grant No. YSBR-062, the Fundamental Research Funds for the Central Universities, the Xiaomi Young Talents Program, and the science research grant from the China Manned Space Project.}

We thank the anonymous referee for their thoughtful suggestions, which have improved this work.


\begin{table*}[h!]
\centering
\caption{Instrument, filter name, effective wavelength in nanometers, width in nanometers, and survey/reference for each filter used in this analysis.}\label{tab:filters}
\begin{tabular}{lcccl}
\hline\hline
\textbf{Instrument} & \textbf{Filter} & \textbf{$\lambda_{eff}$ (nm)} & \textbf{Width (nm)} & \textbf{Survey/Reference} \\ 
\hline
CTIO/DECam & $N419$ & 419.3 & 7.5 & \textbf{ODIN} \citep{odin_survey} \\ 
CTIO/DECam & $N501$ & 501.4 & 7.6 & \textbf{ODIN} \citep{odin_survey} \\ 
CTIO/DECam & $N673$ & 675.0 & 10.0 & \textbf{ODIN} \citep{odin_survey} \\ 
Subaru/Suprime-Cam & $NB711$ & 711.98 & 7.2 & \citet{subaru1, subaru2} \\ 
Subaru/Suprime-Cam & $NB816$ & 815.01 & 11.8 & \citet{subaru1, subaru2} \\ 
\hline
Subaru/Suprime-Cam & $IA427$ & 426.51 & 20.5 & \citet{subaru1, subaru2} \\ 
Subaru/Suprime-Cam & $IA464$ & 463.59 & 21.75 & \citet{subaru1, subaru2} \\ 
Subaru/Suprime-Cam & $IA484$ & 485.0 & 22.8 & \citet{subaru1, subaru2} \\ 
Subaru/Suprime-Cam & $IA505$ & 506.31 & 23.05 & \citet{subaru1, subaru2} \\ 
Subaru/Suprime-Cam & $IA527$ & 526.11 & 24.2 & \citet{subaru1, subaru2} \\ 
Subaru/Suprime-Cam & $IA574$ & 576.48 & 27.25 & \citet{subaru1, subaru2} \\ 
Subaru/Suprime-Cam & $IA624$ & 623.44 & 29.95 & \citet{subaru1, subaru2} \\ 
Subaru/Suprime-Cam & $IA679$ & 678.1 & 33.55 & \citet{subaru1, subaru2} \\ 
Subaru/Suprime-Cam & $IA709$ & 707.36 & 31.6 & \citet{subaru1, subaru2} \\ 
Subaru/Suprime-Cam & $IA738$ & 736.19 & 32.35 & \citet{subaru1, subaru2} \\ 
Subaru/Suprime-Cam & $IA767$ & 768.53 & 36.4 & \citet{subaru1, subaru2} \\ 
Subaru/Suprime-Cam & $IA827$ & 824.13 & 34.05 & \citet{subaru1, subaru2} \\
\hline
HST/WFC3 & $F275W$ & 270.9 & 40.5 & UVCANDELS \citep{UVCANDELS} \\ 
CFHT/MegaCam & $u$ & 378.16 & 72.0 & CFHT-LS \citep{cfht} \\ 
HST/ACS & $F435W$ & 434.13 & 90.3 & UVCANDELS \citep{UVCANDELS} \\ 
Subaru/Suprime-Cam & $B$ & 441.95 & 98.0 & \citet{subaru1, subaru2} \\ 
Subaru/Suprime-Cam & $g^+$ & 474.21 & 133.5 & \citet{subaru1, subaru2} \\ 
CFHT/MegaCam & $g$ & 492.45 & 146.0 & CFHT-LS \citep{cfht} \\ 
Subaru/Suprime-Cam & $V$ & 545.65 & 97.48 & \citet{subaru1, subaru2} \\ 
HST/ACS & $F606W$ & 595.79 & 232.2 & UVCANDELS \citep{UVCANDELS} \\ 
Subaru/Suprime-Cam & $r^+$ & 627.06 & 139.99 & \citet{subaru1, subaru2} \\ 
CFHT/MegaCam & $r$ & 631.62 & 116.0 & CFHT-LS \citep{cfht} \\ 
Subaru/Suprime-Cam & $i^+$ & 795.23 & 136.88 & \citet{subaru1, subaru2} \\ 
HST/ACS & $F814W$ & 807.34 & 182.0 & UVCANDELS \citep{UVCANDELS} \\ 
CFHT/MegaCam & $i$ & 811.84 & 164.0 & CFHT-LS \citep{cfht} \\ 
CFHT/MegaCam & $z$ & 884.68 & 94.0 & CFHT-LS \citep{cfht} \\ 
Subaru/Suprime-Cam & $z^+$ & 906.84 & 117.0 & \citet{subaru1, subaru2} \\ 
VISTA/VIRCAM & $Y$ & 1021.12 & 92.0 & UltraVISTA \citep{ultravista} \\ 
Mayall/NEWFIRM & $J1$ & 1046.88 & 149.0 & NMBS \citep{newfirm} \\ 
Mayall/NEWFIRM & $J2$ & 1195.42 & 149.0 & NMBS \citep{newfirm} \\ 
HST/WFC3 & $F125W$ & 1251.62 & 300.4 & CANDELS \citep{candels1, candels2} \\
VISTA/VIRCAM & $J$ & 1254.11 & 171.0 & UltraVISTA \citep{ultravista} \\ 
Mayall/NEWFIRM & $J3$ & 1278.48 & 140.0 & NMBS \citep{newfirm} \\ 
HST/WFC3 & $F160W$ & 1539.23 & 287.4 & CANDELS \citep{candels1, candels2} \\ 
Mayall/NEWFIRM & $H1$ & 1560.87 & 165.0 & NMBS \citep{newfirm} \\ 
VISTA/VIRCAM & $H$ & 1646.38 & 290.0 & UltraVISTA \citep{ultravista} \\ 
Mayall/NEWFIRM & $H2$ & 1707.21 & 174.0 & NMBS \citep{newfirm} \\ 
VISTA/VIRCAM & $K_s$ & 2148.76 & 308.0 & UltraVISTA \citep{ultravista} \\
Mayall/NEWFIRM & $K$ & 2215.38 & 380.0 & NMBS \citep{newfirm} \\  
Spitzer/IRAC & $ch. 1$ & 3557.26 & 741.08 & S-COSMOS \citep{s-cosmos} \\ 
Spitzer/IRAC & $ch. 2$ & 4504.93 & 1007.22 & S-COSMOS \citep{s-cosmos} \\ 
Spitzer/IRAC & $ch. 3$ & 5738.59 & 1385.04 & S-COSMOS \citep{s-cosmos} \\ 
Spitzer/IRAC & $ch. 4$ & 7927.38 & 2818.16 & S-COSMOS \citep{s-cosmos} \\ 
\hline
\end{tabular}
\end{table*}


\bibliography{sample631}{}

\begin{thebibliography}{}
\expandafter\ifx\csname natexlab\endcsname\relax\def\natexlab#1{#1}\fi
\providecommand{\url}[1]{\href{#1}{#1}}
\providecommand{\dodoi}[1]{doi:~\href{http://doi.org/#1}{\nolinkurl{#1}}}
\providecommand{\doeprint}[1]{\href{http://ascl.net/#1}{\nolinkurl{http://ascl.net/#1}}}
\providecommand{\doarXiv}[1]{\href{https://arxiv.org/abs/#1}{\nolinkurl{https://arxiv.org/abs/#1}}}

\bibitem[{Acquaviva {et~al.}(2012)Acquaviva, Vargas, Gawiser, \& Guaita}]{acquaviva2012curious}
Acquaviva, V., Vargas, C., Gawiser, E., \& Guaita, L. 2012, ApJL, 751, L26, \dodoi{10.1088/2041-8205/751/2/L26}

\bibitem[{Arrabal~Haro {et~al.}(2020)Arrabal~Haro, Rodr{\'\i}guez~Espinosa, Mu{\~n}oz-Tu{\~n}{\'o}n, Sobral, Lumbreras-Calle, Boquien, Hern{\'a}n-Caballero, Rodr{\'\i}guez-Mu{\~n}oz, \& Alcalde~Pampliega}]{arrabal2020differences}
Arrabal~Haro, P., Rodr{\'\i}guez~Espinosa, J., Mu{\~n}oz-Tu{\~n}{\'o}n, C., {et~al.} 2020, MNRAS, 495, 1807, \dodoi{10.1093/mnras/staa1196}

\bibitem[{Ashby {et~al.}(2013)Ashby, Willner, Fazio, Huang, Arendt, Barmby, Barro, Bell, Bouwens, Cattaneo, {et~al.}}]{irac_seds}
Ashby, M., Willner, S., Fazio, G., {et~al.} 2013, ApJ, 769, 80, \dodoi{10.1088/0004-637X/769/1/80}

\bibitem[{Berger \& Zhou(2014)}]{KS_test}
Berger, V.~W., \& Zhou, Y. 2014, Wiley statsref: Statistics reference online, \dodoi{10.1002/9781118445112.stat06558}

\bibitem[{Broussard {et~al.}(2019)Broussard, Gawiser, Iyer, Kurczynski, Somerville, Dav{\'e}, Finkelstein, Jung, \& Pacifici}]{broussard2019star}
Broussard, A., Gawiser, E., Iyer, K., {et~al.} 2019, ApJ, 873, 74, \dodoi{10.3847/1538-4357/ab04ad}

\bibitem[{Calzetti {et~al.}(2000)Calzetti, Armus, Bohlin, Kinney, Koornneef, \& Storchi-Bergmann}]{calzetti2000dust}
Calzetti, D., Armus, L., Bohlin, R.~C., {et~al.} 2000, ApJ, 533, 682, \dodoi{10.1086/308692}

\bibitem[{Carnall {et~al.}(2019)Carnall, Leja, Johnson, McLure, Dunlop, \& Conroy}]{carnall2019measure}
Carnall, A.~C., Leja, J., Johnson, B.~D., {et~al.} 2019, ApJ, 873, 44, \dodoi{10.3847/1538-4357/ab04a2}

\bibitem[{Ceban(2024)}]{ceban2024star}
Ceban, D. 2024, The star formation histories of Lyman-alpha emitters at z= 3 - 7

\bibitem[{Chabrier(2003)}]{chabrier2003galactic}
Chabrier, G. 2003, PASP, 115, 763, \dodoi{10.1086/376392}

\bibitem[{Conroy \& Gunn(2010)}]{FSPS2}
Conroy, C., \& Gunn, J.~E. 2010, ApJ, 712, 833, \dodoi{10.1088/0004-637X/712/2/833}

\bibitem[{Conroy {et~al.}(2009)Conroy, Gunn, \& White}]{FSPS1}
Conroy, C., Gunn, J.~E., \& White, M. 2009, ApJ, 699, 486, \dodoi{10.1088/0004-637X/699/1/486}

\bibitem[{Firestone {et~al.}(2024)Firestone, Gawiser, Ramakrishnan, Lee, Valdes, Park, Yang, Ciardullo, Artale, Benda, {et~al.}}]{odin_lae_sel}
Firestone, N.~M., Gawiser, E., Ramakrishnan, V., {et~al.} 2024, ApJ, 974, 217, \dodoi{10.3847/1538-4357/ad71c9}

\bibitem[{Gawiser {et~al.}(2007)Gawiser, Francke, Lai, Schawinski, Gronwall, Ciardullo, Quadri, Orsi, Barrientos, Blanc, {et~al.}}]{gawiser2007lyalpha}
Gawiser, E., Francke, H., Lai, K., {et~al.} 2007, ApJ, 671, 278, \dodoi{10.1086/522955}

\bibitem[{Grogin {et~al.}(2011)Grogin, Kocevski, Faber, Ferguson, Koekemoer, Riess, Acquaviva, Alexander, Almaini, Ashby, {et~al.}}]{candels1}
Grogin, N.~A., Kocevski, D.~D., Faber, S., {et~al.} 2011, ApJS, 197, 35, \dodoi{10.1088/0067-0049/197/2/35}

\bibitem[{Guaita {et~al.}(2011)Guaita, Acquaviva, Padilla, Gawiser, Bond, Ciardullo, Treister, Kurczynski, Gronwall, Lira, {et~al.}}]{guaita2011lyalpha}
Guaita, L., Acquaviva, V., Padilla, N., {et~al.} 2011, ApJ, 733, 114, \dodoi{10.1088/0004-637X/733/2/114}

\bibitem[{Gwyn(2012)}]{cfht}
Gwyn, S.~D. 2012, ApJ, 143, 38, \dodoi{10.1088/0004-6256/143/2/38}

\bibitem[{Hathi {et~al.}(2016)Hathi, Le~F{\`e}vre, Ilbert, Cassata, Tasca, Lemaux, Le~Brun, Maccagni, Pentericci, {et~al.}}]{hathi2016vimos}
Hathi, N.~P., Le~F{\`e}vre, O., Ilbert, O., {et~al.} 2016, A\&A, 588, A26, \dodoi{10.1051/0004-6361/201526012}

\bibitem[{Heavens {et~al.}(2000)Heavens, Jimenez, \& Lahav}]{MOPED}
Heavens, A.~F., Jimenez, R., \& Lahav, O. 2000, MNRAS, 317, 965, \dodoi{10.1046/j.1365-8711.2000.03692.x}

\bibitem[{Hui \& Gnedin(1997)}]{hui1997equation}
Hui, L., \& Gnedin, N.~Y. 1997, MNRAS, 292, 27, \dodoi{10.48550/arXiv.astro-ph/9612232}

\bibitem[{Iani {et~al.}(2024)Iani, Caputi, Rinaldi, Annunziatella, Boogaard, {\"O}stlin, Costantin, Gillman, P{\'e}rez-Gonz{\'a}lez, Colina, {et~al.}}]{iani2024midis}
Iani, E., Caputi, K.~I., Rinaldi, P., {et~al.} 2024, ApJ, 963, 97, \dodoi{10.3847/1538-4357/ad15f6}

\bibitem[{Im {et~al.}(2024)Im, Hwang, Park, Lee, Song, Appleby, Dubois, Few, Gibson, Kim, {et~al.}}]{im2024testing}
Im, S.~H., Hwang, H.~S., Park, J., {et~al.} 2024, ApJ, 972, 196, \dodoi{10.3847/1538-4357/ad67d2}

\bibitem[{Iyer \& Gawiser(2017)}]{db1-iyer2017}
Iyer, K., \& Gawiser, E. 2017, ApJ, 838, 127, \dodoi{10.3847/1538-4357/aa63f0}

\bibitem[{Iyer(2019)}]{iyer2019reconstructing}
Iyer, K.~G. 2019, PhD thesis, Rutgers The State University of New Jersey, School of Graduate Studies

\bibitem[{Iyer {et~al.}(2019)Iyer, Gawiser, Faber, Ferguson, Kartaltepe, Koekemoer, Pacifici, \& Somerville}]{db2-iyer2019}
Iyer, K.~G., Gawiser, E., Faber, S.~M., {et~al.} 2019, ApJ, 879, 116, \dodoi{10.3847/1538-4357/ab2052}

\bibitem[{Johnson {et~al.}(2021)Johnson, Leja, Conroy, \& Speagle}]{prospector_johnson2021stellar}
Johnson, B.~D., Leja, J., Conroy, C., \& Speagle, J.~S. 2021, ApJS, 254, 22, \dodoi{10.3847/1538-4365/abef67}

\bibitem[{Kikuta {et~al.}(2023)Kikuta, Ouchi, Shibuya, Liang, Umeda, Matsumoto, Shimasaku, Harikane, Ono, Inoue, {et~al.}}]{kikuta2023silverrush}
Kikuta, S., Ouchi, M., Shibuya, T., {et~al.} 2023, ApJS, 268, 24, \dodoi{10.3847/1538-4365/ace4cb}

\bibitem[{Koekemoer {et~al.}(2011)Koekemoer, Faber, Ferguson, Grogin, Kocevski, Koo, Lai, Lotz, Lucas, McGrath, {et~al.}}]{candels2}
Koekemoer, A.~M., Faber, S., Ferguson, H.~C., {et~al.} 2011, ApJS, 197, 36, \dodoi{10.1088/0067-0049/197/2/36}

\bibitem[{Kunth {et~al.}(1998)Kunth, Mas-Hesse, Terlevich, Terlevich, Lequeux, \& Fall}]{kunth1998hst}
Kunth, D., Mas-Hesse, J., Terlevich, E., {et~al.} 1998, A\&A, v. 334, p. 11-20 (1998), 334, 11, \dodoi{10.48550/arXiv.astro-ph/9802253}

\bibitem[{Kusakabe {et~al.}(2018)Kusakabe, Shimasaku, Ouchi, Nakajima, Goto, Hashimoto, Konno, Harikane, Silverman, \& Capak}]{kusakabe2018stellar}
Kusakabe, H., Shimasaku, K., Ouchi, M., {et~al.} 2018, PASJ, 70, 4, \dodoi{10.1093/pasj/psx148}

\bibitem[{Lee {et~al.}(2024)Lee, Gawiser, Park, Yang, Valdes, Lang, Ramakrishnan, Moon, Firestone, Appleby, {et~al.}}]{odin_survey}
Lee, K.-S., Gawiser, E., Park, C., {et~al.} 2024, ApJ, 962, 36, \dodoi{10.3847/1538-4357/ad165e}

\bibitem[{Leja {et~al.}(2019)Leja, Carnall, Johnson, Conroy, \& Speagle}]{dirichlet2}
Leja, J., Carnall, A.~C., Johnson, B.~D., Conroy, C., \& Speagle, J.~S. 2019, ApJ, 876, 3, \dodoi{10.3847/1538-4357/ab133c}

\bibitem[{Leja {et~al.}(2017)Leja, Johnson, Conroy, Van~Dokkum, \& Byler}]{prospector_leja2017deriving}
Leja, J., Johnson, B.~D., Conroy, C., Van~Dokkum, P.~G., \& Byler, N. 2017, ApJ, 837, 170, \dodoi{10.3847/1538-4357/aa5ffe}

\bibitem[{Maier {et~al.}(2003)Maier, Meisenheimer, Thommes, Hippelein, R{\"o}ser, Fried, von Kuhlmann, Phleps, \& Wolf}]{maier2003constraints}
Maier, C., Meisenheimer, K., Thommes, E., {et~al.} 2003, A\&A, 402, 79, \dodoi{10.1051/0004-6361:20030265}

\bibitem[{McCracken {et~al.}(2012)McCracken, Milvang-Jensen, Dunlop, Franx, Fynbo, Le~F{\`e}vre, Holt, Caputi, Goranova, Buitrago, {et~al.}}]{ultravista}
McCracken, H., Milvang-Jensen, B., Dunlop, J., {et~al.} 2012, A\&A, 544, A156, \dodoi{10.1051/0004-6361/201219507}

\bibitem[{Mehta {et~al.}(2024)Mehta, Rafelski, Sunnquist, Teplitz, Scarlata, Wang, Fontana, Hathi, Iyer, Alavi, {et~al.}}]{uvcandels_photozs}
Mehta, V., Rafelski, M., Sunnquist, B., {et~al.} 2024, ApJS, 275, 17, \dodoi{10.3847/1538-4365/ad7d8f}

\bibitem[{Ouchi {et~al.}(2020)Ouchi, Ono, \& Shibuya}]{Ouchi_2020}
Ouchi, M., Ono, Y., \& Shibuya, T. 2020, ARA\&A, 58, 617, \dodoi{10.1146/annurev-astro-032620-021859}

\bibitem[{Papovich {et~al.}(2001)Papovich, Dickinson, \& Ferguson}]{papovich2001stellar}
Papovich, C., Dickinson, M., \& Ferguson, H.~C. 2001, ApJ, 559, 620, \dodoi{10.1086/322412}

\bibitem[{{Partridge} \& {Peebles}(1967)}]{Partridge1967}
{Partridge}, R.~B., \& {Peebles}, P.~J.~E. 1967, \apj, 147, 868, \dodoi{10.1086/149079}

\bibitem[{Ramakrishnan {et~al.}(2023)Ramakrishnan, Moon, Hyeok~Im, Farooq, Lee, Gawiser, Yang, Park, Hwang, Valdes, {et~al.}}]{ramakrishnan2023odin}
Ramakrishnan, V., Moon, B., Hyeok~Im, S., {et~al.} 2023, ApJ, 951, 119, \dodoi{10.3847/1538-4357/acd341}

\bibitem[{Ramakrishnan {et~al.}(2024)Ramakrishnan, Lee, Artale, Yang, Park, Ciardullo, Guaita, Im, Kim, Kumar, {et~al.}}]{ramakrishnan2024odin_protoclusters_and_filaments}
Ramakrishnan, V., Lee, K.-S., Artale, M.~C., {et~al.} 2024, arXiv preprint arXiv:2406.08645, \dodoi{10.48550/arXiv.2406.08645}

\bibitem[{Ramakrishnan {et~al.}(2025)Ramakrishnan, Lee, Firestone, Gawiser, Artale, Gronwall, Guaita, Hwang, Im, Jeong, {et~al.}}]{ramakrishnan2024odin_clustering_of_protoclusters}
Ramakrishnan, V., Lee, K.-S., Firestone, N., {et~al.} 2025, ApJ, 982, 74, \dodoi{10.3847/1538-4357/adb624}

\bibitem[{Reichardt {et~al.}(2001)Reichardt, Jimenez, \& Heavens}]{MOPED_SFH}
Reichardt, C., Jimenez, R., \& Heavens, A.~F. 2001, MNRAS, 327, 849, \dodoi{10.1046/j.1365-8711.2001.04768.x}

\bibitem[{Rhoads {et~al.}(2003)Rhoads, Dey, Malhotra, Stern, Spinrad, Jannuzi, Dawson, Brown, \& Landes}]{rhoads2003spectroscopic}
Rhoads, J.~E., Dey, A., Malhotra, S., {et~al.} 2003, ApJ, 125, 1006, \dodoi{10.1086/346272}

\bibitem[{Rosani {et~al.}(2020)Rosani, Caminha, Caputi, \& Deshmukh}]{rosani2020bright}
Rosani, G., Caminha, G., Caputi, K., \& Deshmukh, S. 2020, A\&A, 633, A159, \dodoi{10.1051/0004-6361/201935782}

\bibitem[{Sanders {et~al.}(2007)Sanders, Salvato, Aussel, Ilbert, Scoville, Surace, Frayer, Sheth, Helou, Brooke, {et~al.}}]{s-cosmos}
Sanders, D.~B., Salvato, M., Aussel, H., {et~al.} 2007, ApJS, 172, 86, \dodoi{10.1086/517885}

\bibitem[{Shapley(2011)}]{shapley2011physical}
Shapley, A.~E. 2011, ARA\&A, 49, 525, \dodoi{10.1146/annurev-astro-081710-102542}

\bibitem[{Simha {et~al.}(2014)Simha, Weinberg, Conroy, Dave, Fardal, Katz, \& Oppenheimer}]{simha2014parametrising}
Simha, V., Weinberg, D.~H., Conroy, C., {et~al.} 2014, arXiv preprint arXiv:1404.0402, \dodoi{10.48550/arXiv.1404.0402}

\bibitem[{Taniguchi {et~al.}(2007)Taniguchi, Scoville, Murayama, Sanders, Mobasher, Aussel, Capak, Ajiki, Miyazaki, Komiyama, {et~al.}}]{subaru1}
Taniguchi, Y., Scoville, N., Murayama, T., {et~al.} 2007, ApJS, 172, 9, \dodoi{10.1086/516596}

\bibitem[{Taniguchi {et~al.}(2015)Taniguchi, Kajisawa, Kobayashi, Shioya, Nagao, Capak, Aussel, Ichikawa, Murayama, Scoville, {et~al.}}]{subaru2}
Taniguchi, Y., Kajisawa, M., Kobayashi, M.~A., {et~al.} 2015, PASJ, 67, 104, \dodoi{10.1093/pasj/psv106}

\bibitem[{Teplitz {et~al.}(2022)Teplitz, Wang, Prichard, Alavi, Rafelski, Grogin, Keokemoer, Sunnquist, \& Mehta}]{https://doi.org/10.17909/8s31-f778}
Teplitz, H., Wang, X., Prichard, L., {et~al.} 2022, Ultraviolet Imaging of the Cosmic Assembly Near-infrared Deep Extragalactic Legacy Survey Fields (UVCANDELS),  STScI/MAST, \dodoi{10.17909/8S31-F778}

\bibitem[{Thommes \& Meisenheimer(2005)}]{thommes2005expected}
Thommes, E., \& Meisenheimer, K. 2005, A\&A, 430, 877, \dodoi{10.1051/0004-6361:20035863}

\bibitem[{Tojeiro {et~al.}(2007)Tojeiro, Heavens, Jimenez, \& Panter}]{VESPA_SFH}
Tojeiro, R., Heavens, A.~F., Jimenez, R., \& Panter, B. 2007, MNRAS, 381, 1252, \dodoi{10.1111/j.1365-2966.2007.12323.x}

\bibitem[{Vargas {et~al.}(2014)Vargas, Bish, Acquaviva, Gawiser, Finkelstein, Ciardullo, Ashby, Feldmeier, Ferguson, Gronwall, {et~al.}}]{vargas2014stack}
Vargas, C.~J., Bish, H., Acquaviva, V., {et~al.} 2014, ApJ, 783, 26, \dodoi{10.1088/0004-637X/783/1/26}

\bibitem[{Venemans {et~al.}(2005)Venemans, R{\"o}ttgering, Miley, Kurk, De~Breuck, Overzier, van Breugel, Carilli, Ford, Heckman, {et~al.}}]{venemans2005properties}
Venemans, B.~P., R{\"o}ttgering, H., Miley, G., {et~al.} 2005, A\&A, 431, 793, \dodoi{10.1051/0004-6361:20042038}

\bibitem[{Wang {et~al.}(2025)Wang, Teplitz, Smith, Windhorst, Rafelski, Mehta, Alavi, Ji, Brammer, Colbert, {et~al.}}]{UVCANDELS}
Wang, X., Teplitz, H.~I., Smith, B.~M., {et~al.} 2025, ApJ, 980, 74, \dodoi{10.3847/1538-4357/ada4ab}

\bibitem[{Whitaker {et~al.}(2011)Whitaker, Labb{\'e}, van Dokkum, Brammer, Kriek, Marchesini, Quadri, Franx, Muzzin, Williams, {et~al.}}]{newfirm}
Whitaker, K.~E., Labb{\'e}, I., van Dokkum, P.~G., {et~al.} 2011, ApJ, 735, 86, \dodoi{10.1088/0004-637X/735/2/86}

\bibitem[{White {et~al.}(2024)White, Raichoor, Dey, Garrison, Gawiser, Lang, Lee, Myers, Schlegel, Valdes, {et~al.}}]{white2024clustering}
White, M., Raichoor, A., Dey, A., {et~al.} 2024, JCAP, 2024, 020, \dodoi{10.1088/1475-7516/2024/08/020}

\end{thebibliography}
\bibliographystyle{aasjournal}

\end{document}